# Diffraction of thermal radiation from binary anisotropic structure

Nir Dahan, Yuri Gorodetski, Kobi Frischwasser, Vladimir Kleiner and Erez Hasman

Micro and Nanooptics Laboratory, Faculty of Mechanical Engineering, and Russell Berrie Nanotechnology Institute, Technion – Israel Institute of Technology, Haifa 32000, Israel

**Abstract**

Thermal emission from binary grating on SiC wafer supported by phonon-polaritons is analyzed. The structure is comprised of homogeneous grating domains, whose orientation is parallel and perpendicular to the *x*-axis. The dispersion relation of the emitted light corresponds to translation symmetry of the structure.

Thermal emission has been shown to be modified by utilizing the high density of states of surface waves and their long-range propagation[1,2]. The coupling of non-radiative surface modes to radiative modes can be achieved by performing a periodic perturbation on the surface, which provides a momentum-matching that produces a coherent and polarized emission[2-6].

We start from investigation of the superstructure illustrated in the inset in Fig. 1a. The superstructure domains with a local periodicity $\Lambda = 11.6 \mu m$ and a depth of 300nm were realized using standard photolithographic techniques on a SiC substrate which supports surface phonon polariton (SPhP) in the infrared region.

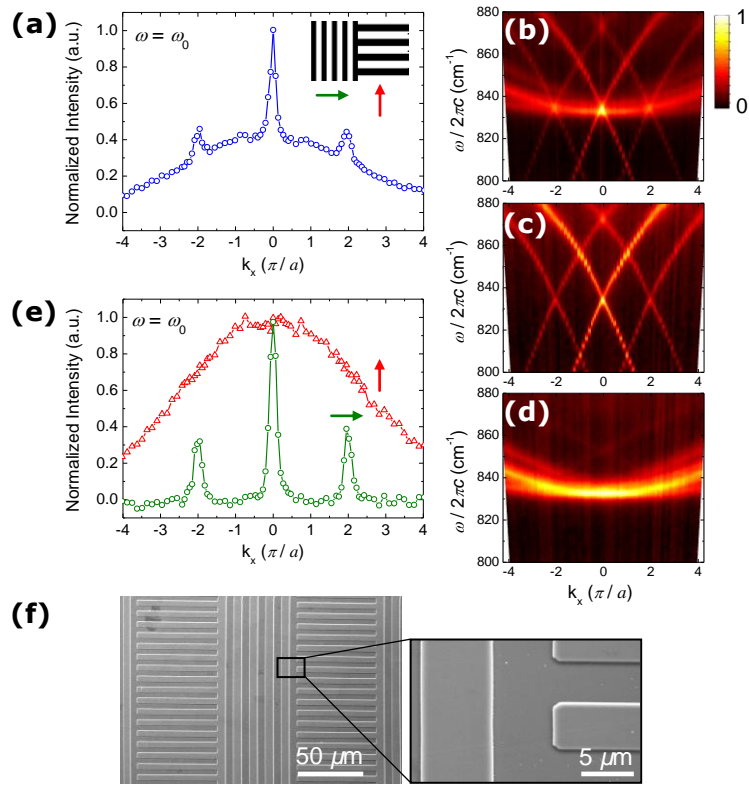

**Figure 1. Emission of superstructure illustrated in the inset in panel a: a**, Angular emission at $\omega_0$ without polarizer. **b**, Dispersion relation measured without polarizer. **c**, Dispersion relation measured with polarizer parallel to the principal direction – $x$ axis. **d**, Dispersion relation measured with polarizer perpendicular to the principal direction – $y$ axis. **e**, Cross sections of panel c (green circle) and panel d (red triangle) at $\omega_0$. **f**, SEM images of the superstructure.

Figure 1b shows the dispersive emission measured from the superstructure with a cross section at $\omega_0$ in Fig. 1a ($\omega_0/2\pi c = 833.6 \text{cm}^{-1}$ - corresponding to $\lambda_0 = 12\mu\text{m}$). As can be seen in Fig. 1b, the signature of the SPhP dispersion observed from the homogeneous structure is clearly seen in the center, in addition to the two SPhP dispersion curves centered at spatial frequencies $\pm 2\pi/a$ where $a$ is the distance along the *x* axis for the $\pi$- rotation ($a = 140\mu m$). Moreover, an omnidirectional emission appears at the resonant frequency. To comprehend the origin of this dispersion, we measured the emission with a polarizer parallel (Fig. 1c) and perpendicular (Fig. 1d) to the principal axis – the superstructure's wave vector. The cross sections at $\omega_0$ are plotted in Fig. 1e. We can clearly see that the emission given in Fig. 1b is a superposition of these two polarization states. The polarization along the *x* axis (green circles) corresponds to domains with a local grating wave vector that is parallel to the superstructure's wave vector. Therefore, diffraction from domain arrays with a period *a* is observed as anticipated from a structure with a translation symmetry,

$$\omega = \omega(k_x + mG), \qquad (1)$$

Furthermore, the lobe width shown in Fig. 1e is comparable to the lobe width of the homogeneous structure (not shown) associated with the spatial coherence length of SPhPs $l_c = 1/\text{Im}(k_{SPP})$. To the contrary, polarized emission along the *y* axis (red triangles) exhibits a diffraction with a lobe width that corresponds to an aperture of *a*/2 associated with the domain size. Hence, the structure behaves like a weakly coupled array of thermal sources resulting in a slow mode.

**Polarization analysis**:

The Stokes parameters are determined by measuring the intensity of the radiative field through different combinations of a polarizer and retarder. In our experiment, we used a circular polarizer, i.e., a wave plate followed by a polarizer (see Fig. 2) in different orientations. Usually, this measurement is for monochromatic light where the retardation of the wave plate is given for a particular wavelength. For a broad spectrum of light, as in thermal emission, this retardation value is no longer valid.

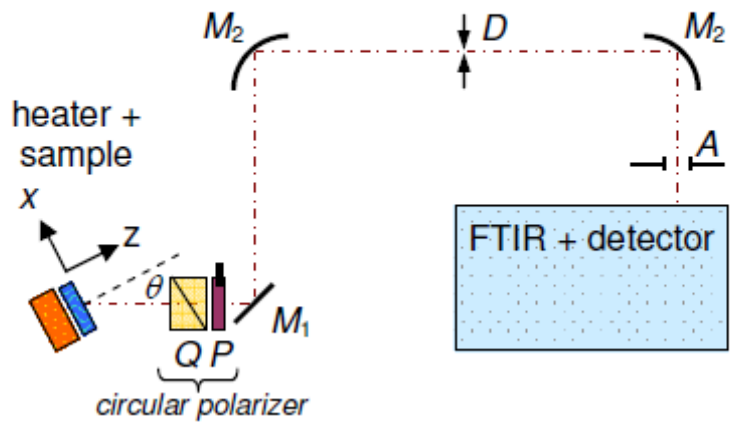

**Figure 2. Experimental set-up:** measurement of spectral emission at angle $\theta$. $P$ – polarizer; $Q$ – quarter-wave plate for 10.6μm wavelength; $M_1$ – flat mirror on a rotating stage; $M_2$ – parabolic mirror, focal length = 250mm; $D$ – angular resolution slit in the focal plane of $M_2$, width = 1mm; $A$ – field of view aperture, diameter = 8mm; FTIR – Fourier transform infrared spectrometer.

Therefore, we assume retardation that is wavelength dependent as

$\tilde{\phi}(\lambda) = \frac{2\pi}{\lambda}(n_o - n_e)d$, where $n_o$ ($n_e$) is the ordinary (extraordinary) refractive index.

If we know that the retardation in a quarter wave plate follows $\tilde{\phi}' = \frac{2\pi}{\lambda'}(n_o - n_e)d$ for

$\lambda' = 10.6\mu$m wavelength, $d$ is the wave plate thickness, and $(n_o - n_e)$ is approximately

constant in our spectral range, we can use the retardation as $\tilde{\phi}(\lambda) \approx \frac{\lambda'}{\lambda}\tilde{\phi}'$. In the case of a quarter wave plate, $\tilde{\phi}(\lambda) \approx \frac{\pi}{2}\frac{\lambda'}{\lambda}$. Let $\psi$ be the retarder orientation and $\vartheta$ be the polarizer transmission axis, the intensity can be calculated using the Mueller matrix as,

$$I(\psi,\vartheta,\tilde{\phi}) = \frac{S_0}{2} +$$
$$\frac{S_1}{2}\left[\cos 2\psi(\cos 2\vartheta \cdot \cos 2\psi + \sin 2\vartheta \cdot \sin 2\psi) + \sin 2\psi \cdot \cos\tilde{\phi}(\sin 2\psi - \cos 2\psi)\right] +$$
$$\frac{S_2}{2}\left[\sin 2\psi(\cos 2\vartheta \cdot \cos 2\psi + \sin 2\psi) + \cos 2\psi \cdot \cos\tilde{\phi}(\sin 2\vartheta \cdot \cos 2\psi - \sin 2\psi)\right] +$$
$$\frac{S_3}{2}\left[\sin\tilde{\phi}(\cos 2\vartheta \cdot \sin 2\psi - \sin 2\vartheta \cdot \cos 2\psi)\right]$$
(2)

The intensities were captured at different orientations of the retarder and polarizer for which $\psi = 0, 45°, -45°, 45°$ and $\vartheta = 0, 0, 0, 45°$, respectively. With these four combinations, we find the Stokes parameters as a function of the different intensities to be

$$S_0 = \left(I(45°,0°) + I(-45°,0°) - 2I(0°,0°)\cos\tilde{\phi}\right)/\left(1 - \cos\tilde{\phi}\right), \quad (3)$$

$$S_1 = 2I(0°,0°) - S_0, \quad (4)$$

$$S_2 = 2I(45°,45°) - S_0, \quad (5)$$

$$S_3 = \left(I(45°,0°) - I(-45°,0°)\right)/\sin\tilde{\phi}. \quad (6)$$

**Methods:**

*Fabrication* – The SiC (6H-polytype) samples were fabricated by standard photolithographic techniques using a negative photoresist. After developing the photoresist, a 1300 Å thick layer of NiCr was deposited and then a lift-off was performed. The substrate was etched through the NiCr mask by reactive ion etching at a power of 250W and a pressure of 10 mTorr with $SF_6$ and $O_2$ gases at flow rates of 19 sccm and 1 sccm, respectively. The etching was performed at a rate of 600 Å/minute at room temperature for 5 minutes. As a final step, the remaining NiCr was removed with a Cr etchant.

*Experimental set-up* – The spectral and directional emissivity were measured by mirror optics Fig. 2. The sample was heated to $770 \pm 1K$. The sample temperature was measured with a K-type thermocouple and controlled by a temperature controller (HeatWave Labs, model 101303-04B). The measurements of the emission spectra in the range of angles $\theta$, $[-50^0 \div 50^0]$ were performed using a Fourier transform infrared spectrometer (Bruker − Vertex 70) equipped with a cooled HgCdTe detector. The spectral resolution was set to 1cm$^{-1}$, the field of view was chosen as 8mm to avoid edge effects (each sample is 12mm square), and an angular resolution 0.1°.